\documentclass{WileyMSP-template}

\usepackage[dvipsnames]{xcolor}

\newcommand{\eg}{e.\,g.}
\newcommand{\ie}{i.\,e.}

\newcommand{\rev}[1]{\textcolor{Black}{#1}}

\begin{document}

\pagestyle{fancy}
\rhead{\includegraphics[width=2.5cm]{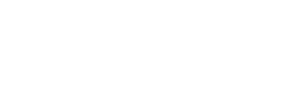}}

\title{Artificial Intelligence-based Clinical Decision Support\\for COVID-19 -- Where Art Thou?}

\maketitle


\author{Mathias Unberath*}
\author{Kimia Ghobadi}
\author{Scott Levin}
\author{Jeremiah Hinson}
\author{Gregory D Hager}


\dedication{To all front line health workers.}

\begin{affiliations}
The Malone Center for Engineering in Healthcare\\
Johns Hopkins University\\
3400 N Charles Street, Malone Hall Suite 340\\
$\{$mathias, kimia, slevin33, hinson, hager$\}$@jhu.edu\\

All authors have contributed equally to this manuscript.
\end{affiliations}


\keywords{COVID-19, Artificial Intelligence, Computer Assisted Diagnosis, Prognosis, Data Science, Healthcare}

\begin{abstract}

The COVID-19 crisis has brought about new clinical questions, new workflows, and accelerated distributed healthcare needs. While artificial intelligence (AI)-based clinical decision support seemed to have matured, the application of AI-based tools for COVID-19 has been limited to date. In this perspective piece, we identify opportunities and requirements for AI-based clinical decision support systems and highlight challenges that impact ``AI readiness" for rapidly emergent healthcare challenges. 

\end{abstract}

\section{Background}
Prior to January 2020, the artificial intelligence and machine learning (AI/ML) for healthcare community had many reasons to be pleased with the recent progress of their field. Learning-based algorithms had been shown to accurately forecast the onset of septic shock~\cite{henry2015targeted}, ML-based pattern recognition methods classified skin lesions with dermatologist level accuracy~\cite{esteva2017dermatologist}, diagnostic AI systems successfully identified diabetic retinopathy during routine primary care visits~\cite{abramoff2018pivotal}, AI-based breast cancer screening outperformed radiologists by a fairly large margin~\cite{mckinney2020international}, ML-driven triaging tools improved outcome differentiation beyond the emergency severity index~\cite{levin2018machine}, AI-enabled assistance systems simplified interventional workflows~\cite{vercauteren2019cai4cai}, and algorithm-driven organizational studies enabled redesign of infusion centers~\cite{ghobadi2017translating}. Many would have argued that, after nearly 60 years on the test bench~\cite{yu2018artificial}, AI in healthcare had finally reached a level of maturity, performance, and reliability that was compatible with the unforgiving requirements imposed by clinical practice. 

Today, only a few months later, this rather sunny outlook has become overcast. The world’s healthcare systems are facing the outbreak of a novel respiratory disease, COVID-19. \rev{As of May 29, 2020 more than 5,800,000 cases of COVID-19 -- caused by severe acute respiratory syndrome coronavirus 2 (SARS-CoV-2) infection -- have been reported across 210 countries, leading to more than 360,000 deaths~\cite{JHU_CSSE_Dashboard}.} Although a rapidly emerging infectious disease escalating to a global pandemic represents a rare worst-case scenario, it provides an opportunity to observe the resilience of each and every link in the healthcare chain. And, though heavily stressed, most links of this chain withstood the stress test of this first wave of the pandemic
, while the highly acclaimed AI/ML link appeared to have given in. This breakdown, however, cannot be attributed to a lack of opportunity -- on the contrary: Several articles in popular and scientific press describe immediate opportunities for AI-assisted tools to have a positive impact on patient outcomes. These applications range from improved diagnosis, triaging and prognostication~\cite{jiang2020towards,jiang2020review} to personalized treatment decision support~\cite{van2020artificial} to automated monitoring tools~\cite{Strickland_2020_IEEESpectrum}. These AI-based clinical decision support (CDS) are the central focus of this perspective piece. 

Why, then, are AI-assisted CDS tools seemingly limited in contributing to the fight against COVID-19?

\section{Identifying Opportunities}
\label{sec:question}

While the COVID-19 crisis has brought about new clinical questions, new workflows, and accelerated distributed healthcare needs, the core principles of successful CDS are unwavering~\cite{kawamoto2005improving} -- indeed they are arguably more important than ever. Foremost, an AI-based CDS must address a clinical decision where variability and uncertainty exists, since achieving reproducibility constitutes a core challenge in the course of the diagnosis or treatment continuum~\cite{lindsell2020action}. Many opportunities and unknowns exist in both identifying patients with SARS-CoV-2 and performing optimal treatment while containing transmission. 

Clinical needs and opportunities for AI-driven CDS are evolving with the pandemic. Since early on, especially in areas with limited diagnostic testing capacity (including most of the United States), rapid identification of patients with COVID-19 has been a major challenge. New signs and symptoms of the illness are still being described, and guidance on which patients should be considered persons under investigation (PUIs) for COVID-19 or tested for SARS-CoV-2 infection is continuously evolving~~\cite{centers2020evaluating}. Failure to rapidly identify and isolate patients with COVID-19 in the hospital setting creates opportunity for nosocomial infection spread, placing patients and healthcare workers at excess risk. Screening and isolation procedures that lack specificity drive over-use of limited \rev{personal protective equipment (PPE)} and delays to care for non-COVID-related illness. To date, our approach to screening for COVID-19 has been largely reactive. AI-driven algorithms with capacity to identify aberrant clinical presentations or to quickly identify non-obvious patterns in patients with the disease could shift this approach to a more proactive one. Beyond screening, identification of individual patient factors (\eg, age and comorbidities), symptoms and clinical findings or measures (\eg, vital signs, laboratory and imaging results) that are predictive of clinical deterioration is needed. AI-driven CDS could be used to target earlier intervention or guide disposition (\eg, discharge versus hospitalization) decision-making based on these predictions. As the pandemic continues to unfold, clinical needs will also change. For example, AI-driven CDS to support optimal resource allocation, load balancing, or treatment selection may become the predominant need. 

While the application of AI-based CDS for COVID-19 has been limited to date, this is somewhat understandable. AI is highly dependent on sufficiently diverse, representative data; AI/ML algorithms cannot be trained and adequately validated without creating large data sets that reflect the clinical use case, and such data will always be limited in the early phases of an emerging infectious disease. Nonetheless, the need and opportunity for improvement of clinical care through AI will continue to increase over the coming months -- and those who are well-positioned to accrue and capitalize on data will be able to seize this opportunity.

In addition to capturing data, it is equally important to understand the clinical use case -- effective development and deployment of any CDS, including that driven by AI, requires a deep understanding of both the problems of focus and the environment in which it is encountered. For instance, an AI-driven prediction model is effective only when its prediction can be linked to a clear action (\eg ~altered clinical decision-making that prevents or better prepares for the predicted outcome) an is embedded seamlessly in the relevant clinical workflow. Rapid identification of the most pressing clinical needs, and understanding which of the many targets for AI that will yield improvements in clinical care and effective development of CDS requires multi-disciplinary (clinician-engineer) collaboration. Teams with existing collaborations of this nature, such as in our own Malone Center for Engineering in Healthcare and Center for Data Science in Emergency Medicine, will be best positioned to rapidly respond to and address needs as they emerge~\cite{lindsell2020action}.

\section{Capitalizing on Data}

As already noted, AI algorithms cannot be trained and validated without sufficiently large quantities of data, but at the same time not all data that may be available is necessarily useful. AI-based CDS tools must be designed with a target population in mind that is determined by the clinical question to be answered. To obtain unbiased estimates of the CDS system’s performance in deployment, careful study design is an essential pre-requisite. When curating datasets for model development and validation, the in- and exclusion criteria for any sample must be well-defined and precisely aligned with the patient cohort who would later be subject to algorithmic analysis. Collecting data in this manner is time consuming and requires care and effort, which is especially challenging during a rapidly evolving pandemic situation. It is not surprising that many of the recent manuscripts that describe AI-based CDS tools rely on \emph{convenience sampling} of data, \ie\ dataset curation strategies that prioritize easily available data, \eg, public repositories such as challenges or The Cancer Imaging Archive~\cite{clark2013cancer}, without considering the ultimate target population. From an engineering perspective, such studies can confirm the validity of the technical approach, \emph{but they cannot provide evidence to support immediate clinical deployment}. 

Consequently, while there are early reports on the use of AI-based prediction models for detection and prognosis of COVID-19, the reported results have a high risk of bias and are probably overly optimistic~\cite{wynants2020prediction}. It is paramount that these promising technical feasibility studies are followed up with carefully designed validation studies to demonstrate the model’s usefulness in the target population. The challenges that this creates are not new. They have already been recognized more than two decades ago by Jerome Kassirer~\cite{kassirer1994report} who concludes: ``\emph{Typically, new models were tested, found to exhibit interesting characteristics, and then abandoned when their developers found that the next steps were far harder than designing the first model}''. As highlighted in Section~\ref{sec:question}, CDS tools are urgently needed and careful reporting of the study and dataset design, \eg\ following TRIPOD~\cite{collins2015transparent}, is critical to interpret the clinical implications of the proposed CDS tool.

Irrespective of the sampling strategy, data collection for AI-based CDS development involves the aggregation and pre-conditioning of large amounts of medical records. Information in patient medical records is considered highly sensitive and private information that must be protected -- these standards are governed by the well-known HIPAA privacy act. Because research using identifiable protected health information qualifies as human subjects research, any such study must be approved by the respective Institutional Review Board (IRB), which are established to protect the welfare, rights, and privacy of human subjects. While research on fully anonymized patient data does not qualify as human subjects research and is thus exempt, this determination cannot usually be made by the study team but is reserved for the IRB.

A major challenge for all data-driven investigations is the growing evidence that the technical feasibility of fully automated anonymization~\cite{benitez2010evaluating,el2011systematic} is questionable. This in turn complicates a straight-forward ``exempt'' declaration, suggesting that new study protocols will have to undergo full IRB consideration. If the IRB is operating at capacity, it can take multiple weeks before a final decision is rendered, and if positive, data collection can begin. Further, in light of the COVID-19 state of emergency, many new studies are being submitted for review and IRBs must prioritize protocols that promise direct and immediate benefits to participants, such as controlled trials of emerging treatments~\cite{wang2020remdesivir,casadevall2020convalescent}. This circumstance may further delay the approval of studies that have low immediate benefit to the subjects, which is the case for most AI algorithm development. Consequently, even if predictor and outcome variables are known immediately after onset, thus enabling multi-disciplinary teams to define value-creating AI use cases, the hands-on work towards this goal will likely be delayed substantially due to IRB and data trust related processes. This circumstance seems to be supported by observing the delay between the onset of the epidemic in late December 2019 and the number of publications describing ML algorithms for the automated interpretation of imaging data that only surged in late April 2020. 

While the authors unanimously advocate for the importance of IRB and data trust review of studies to protect human subjects, it is becoming increasingly clear that current practices and processes do not easily support the rapid development and deployment of AI technology. To promote \emph{AI readiness}, we should consider the implementation of IRB sub-committees dedicated to AI algorithm development with specialized study protocol templates that allow for rapid turnaround review of data science-related projects. If indeed the implementation of AI and evidence-based, personalized medicine in routine clinical care is an institutional priority, then clear guidelines and pre-approved mechanisms should be created that empower teams to move quickly and have impact whilst, by design, complying with all ethical standards.

\section{Infrastructure}
Combating COVID-19 has been challenging both due to the large number of asymptomatic patients who can spread the virus and the highly variable impact of SARS-CoV-2 on different patients. The questions that need answering range from identifying and testing patients, therapeutic treatments, to protecting healthcare workers, and, perhaps more challenging, predicting and responding to the healthcare need which triggered the ''flatten the curve'' response and the widespread shutdowns to keep the healthcare systems afloat. While reducing physical distance is an effective way to curb the sudden impact on healthcare systems, it comes at a large economical and societal cost. Alternative approaches, however, require a thorough understanding of the impacts of COVID-19 on healthcare systems operations, diagnoses, and treatments, as well as devising smart testing and tracing responses. This approach requires proactive actions that are informed through a robust data infrastructure. It is overly simplistic to assume that a global pandemic could be controlled without proper use of the growing amount of data and knowledge that is generated by the minute. At the same time, large and fast-paced data and knowledge aggregations cannot be processed without the help of AI, which in return, is empowered by a reliable data infrastructure that is accessible to researchers in a timely manner. This access should be significantly expedited by establishing robust pathways for data sharing and usage, including the standardization of data formats.  

To enable such an infrastructure, multiple data components are needed. Consider for instance, a dataset on hospital operations that summarizes basic information such as employees shift work or the number of utilized resources, including PPEs -- As trivial as this data sounds, it has proven a challenge to gather this data in a reliable manner, and lack of such data is ever more felt when a crisis like the current pandemic hits and the need for fast load-balancing and resource allocation rise seemingly overnight. Another important component of any data infrastructure is the availability of clinical data in a standard form that can be shared in an anonymized manner across healthcare systems. This data is particularly important in informing diagnosis and response to therapy -- one of the early challenges that clinicians faced during the COVID-19 pandemic. While some systems exist that capture similar data, they are often incomplete and silo-ed to the particular organization that generated the data, with accompanying challenges including IRB approvals. Rarely can such datasets be shared on a large scale across multiple systems, although there are noteworthy exceptions~\cite{ResearchDatabase}. An important differential characteristic of an ideal data infrastructure and a key point that enables healthcare systems to respond quickly and proactively, is that such data infrastructure should be devised with data sciences and AI in mind, as opposed to considering it as an afterthought to what organically happens in healthcare environments. An infrastructure that is multi-disciplinary and is informed by not only the clinicians and medical researchers, but also data scientists, engineers, managerial sciences, and public health experts is the one that has a better chance to stand the test of time.

\section{Looking Ahead}
As the pandemic progresses and as more data is acquired, becomes available for research, and is shared, it is likely that AI’s contribution to combating the COVID-19 crisis will grow. In non-human subjects research tasks, such as language processing or computational biology, AI has already started to make an impact by extracting key findings from the ever increasing body of literature on COVID-19~\cite{KaggleCovid} or by trying to understand the protein structure of SARS-CoV-2 to drive drug discovery~\cite{senior2020improved}.
 
\rev{When considering AI/ML for healthcare, we note a number of recent developments that seek to address some of the roadblocks identified herein. Commendable examples include FDA's request for feedback on their proposed regulatory framework for AI/ML-based software as medical device~\cite{us2019proposed}, and the Office for Human Research Protections' Exploratory Workshop Series on privacy and health research, which discusses potential operational solutions to the challenges IRBs face when reviewing ''big data research''~\cite{HHS_workshop}. Specifically for CDS, AI’s notable absence in the early phases of this initial surge will likely not be the whole story. There are pockets of examples~\cite{Thirona,StoCastiNews} where AI is being used to support COVID-19 related patient care right now, and many more examples will follow.}

Once the dust of this first wave of the pandemic has settled, we, as a community, should spend some time to identify and analyze the organizational, institutional, or regulatory hurdles that this healthcare crisis has highlighted, as well as the solution paths that emerged to bring AI-based CDS systems for COVID-19 to the bedside. These insights should contribute to an open discussion of how we can improve on the \emph{AI readiness} of current practices and protocols. After all, who knows when the next severe test comes forth and what it will look like. Let's be prepared!

\medskip
\textbf{Acknowledgements} \par 
This article was supported by institutional funds provided by the Malone Center for Engineering in Healthcare.

\medskip

%

\textbf{References}\\

\begin{thebibliography}{10}
	\providecommand{\url}[1]{\texttt{#1}}
	\providecommand{\urlprefix}{URL }
	
	\bibitem{henry2015targeted}
	K.~E. Henry, D.~N. Hager, P.~J. Pronovost, S.~Saria,
	\newblock \emph{Science translational medicine} \textbf{2015}, \emph{7}, 299
	299ra122.
	
	\bibitem{esteva2017dermatologist}
	A.~Esteva, B.~Kuprel, R.~A. Novoa, J.~Ko, S.~M. Swetter, H.~M. Blau, S.~Thrun,
	\newblock \emph{Nature} \textbf{2017}, \emph{542}, 7639 115.
	
	\bibitem{abramoff2018pivotal}
	M.~D. Abr{\`a}moff, P.~T. Lavin, M.~Birch, N.~Shah, J.~C. Folk,
	\newblock \emph{NPJ digital medicine} \textbf{2018}, \emph{1}, 1 1.
	
	\bibitem{mckinney2020international}
	S.~M. McKinney, M.~Sieniek, V.~Godbole, J.~Godwin, N.~Antropova, H.~Ashrafian,
	T.~Back, M.~Chesus, G.~C. Corrado, A.~Darzi, et~al.,
	\newblock \emph{Nature} \textbf{2020}, \emph{577}, 7788 89.
	
	\bibitem{levin2018machine}
	S.~Levin, M.~Toerper, E.~Hamrock, J.~S. Hinson, S.~Barnes, H.~Gardner,
	A.~Dugas, B.~Linton, T.~Kirsch, G.~Kelen,
	\newblock \emph{Annals of emergency medicine} \textbf{2018}, \emph{71}, 5 565.
	
	\bibitem{vercauteren2019cai4cai}
	T.~Vercauteren, M.~Unberath, N.~Padoy, N.~Navab,
	\newblock \emph{Proceedings of the IEEE} \textbf{2019}, \emph{108}, 1 198.
	
	\bibitem{ghobadi2017translating}
	K.~Ghobadi, A.~C. Zenteno, A.~R. Marshall, P.~F. Dunn, R.~Levi, J.~H. Stone,
	\newblock \emph{NEJM Catalyst} \textbf{2017}, \emph{3}, 1.
	
	\bibitem{yu2018artificial}
	K.-H. Yu, A.~L. Beam, I.~S. Kohane,
	\newblock \emph{Nature biomedical engineering} \textbf{2018}, \emph{2}, 10 719.
	
	\bibitem{JHU_CSSE_Dashboard}
	{Johns Hopkins University - Center for Systems Science and Engineering (CSSE)
		COVID-19 Dashboard},
	\newblock \url{https://coronavirus.jhu.edu/map.html},
	\newblock Accessed: 2020-05-29.
	
	\bibitem{jiang2020towards}
	X.~Jiang, M.~Coffee, A.~Bari, J.~Wang, X.~Jiang, J.~Huang, J.~Shi, J.~Dai,
	J.~Cai, T.~Zhang, et~al.,
	\newblock \emph{CMC-Computers, Materials \& Continua} \textbf{2020}, \emph{63},
	1 537.
	
	\bibitem{jiang2020review}
	F.~Jiang, L.~Deng, L.~Zhang, Y.~Cai, C.~W. Cheung, Z.~Xia,
	\newblock \emph{Journal of General Internal Medicine} \textbf{2020}, 1--5.
	
	\bibitem{van2020artificial}
	M.~van~der Schaar, A.~Alaa \textbf{2020}.
	
	\bibitem{Strickland_2020_IEEESpectrum}
	{IEEE Spectrum -- AI Can Help Hospitals Triage COVID-19 Patients},
	\newblock
	\url{https://spectrum.ieee.org/the-human-os/artificial-intelligence/medical-ai/ai-can-help-hospitals-triage-covid19-patients},
	\newblock Accessed: 2020-04-30.
	
	\bibitem{kawamoto2005improving}
	K.~Kawamoto, C.~A. Houlihan, E.~A. Balas, D.~F. Lobach,
	\newblock \emph{Bmj} \textbf{2005}, \emph{330}, 7494 765.
	
	\bibitem{lindsell2020action}
	C.~J. Lindsell, W.~W. Stead, K.~B. Johnson,
	\newblock \emph{JAMA} \textbf{2020}.
	
	\bibitem{centers2020evaluating}
	C.~for Disease~Control, Prevention, et~al.,
	\newblock \emph{Centers for Disease Control and Prevention. Updated}
	\textbf{2020}.
	
	\bibitem{clark2013cancer}
	K.~Clark, B.~Vendt, K.~Smith, J.~Freymann, J.~Kirby, P.~Koppel, S.~Moore,
	S.~Phillips, D.~Maffitt, M.~Pringle, et~al.,
	\newblock \emph{Journal of digital imaging} \textbf{2013}, \emph{26}, 6 1045.
	
	\bibitem{wynants2020prediction}
	L.~Wynants, B.~Van~Calster, M.~M. Bonten, G.~S. Collins, T.~P. Debray,
	M.~De~Vos, M.~C. Haller, G.~Heinze, K.~G. Moons, R.~D. Riley, et~al.,
	\newblock \emph{bmj} \textbf{2020}, \emph{369}.
	
	\bibitem{kassirer1994report}
	J.~P. Kassirer,
	\newblock A report card on computer-assisted diagnosis--the grade: C,
	\textbf{1994}.
	
	\bibitem{collins2015transparent}
	G.~S. Collins, J.~B. Reitsma, D.~G. Altman, K.~G. Moons,
	\newblock \emph{British Journal of Surgery} \textbf{2015}, \emph{102}, 3 148.
	
	\bibitem{benitez2010evaluating}
	K.~Benitez, B.~Malin,
	\newblock \emph{Journal of the American Medical Informatics Association}
	\textbf{2010}, \emph{17}, 2 169.
	
	\bibitem{el2011systematic}
	K.~El~Emam, E.~Jonker, B.~M. Luk~Arbuckle,
	\newblock \emph{PloS one} \textbf{2011}, \emph{6}, 12.
	
	\bibitem{wang2020remdesivir}
	Y.~Wang, D.~Zhang, G.~Du, R.~Du, J.~Zhao, Y.~Jin, S.~Fu, L.~Gao, Z.~Cheng,
	Q.~Lu, et~al.,
	\newblock \emph{The Lancet} \textbf{2020}.
	
	\bibitem{casadevall2020convalescent}
	A.~Casadevall, L.-a. Pirofski, et~al.,
	\newblock \emph{The Journal of clinical investigation} \textbf{2020},
	\emph{130}, 4 1545.
	
	\bibitem{ResearchDatabase}
	{COVID-19 Research Database},
	\newblock \url{https://covid19researchdatabase.org/},
	\newblock Accessed: 2020-05-14.
	
	\bibitem{KaggleCovid}
	{Kaggle - COVID-19 Kaggle community contributions},
	\newblock \url{https://www.kaggle.com/covid-19-contributions},
	\newblock Accessed: 2020-05-14.
	
	\bibitem{senior2020improved}
	A.~W. Senior, R.~Evans, J.~Jumper, J.~Kirkpatrick, L.~Sifre, T.~Green, C.~Qin,
	A.~{\v{Z}}{\'\i}dek, A.~W. Nelson, A.~Bridgland, et~al.,
	\newblock \emph{Nature} \textbf{2020}, 1--5.
	
	\bibitem{us2019proposed}
	{US Food and Drug Administration}, et~al.,
	\newblock Proposed regulatory framework for modifications to artificial
	intelligence/machine learning (ai/ml)-based software as a medical device
	(samd)—discussion paper and request for feedback. 2019, \textbf{2019}.
	
	\bibitem{HHS_workshop}
	{US Department of Health \& Human Services: 2019 Office for Human Research
		Protection Exploratory Workshop -- Privacy and Health Research in a
		Data-Driven World},
	\newblock
	\url{https://www.hhs.gov/ohrp/education-and-outreach/exploratory-workshop/2019-workshop/index.html},
	\newblock Accessed: 2020-05-29.
	
	\bibitem{Thirona}
	{Thirona - Artificial intelligence to Screen for COVID-19 on CT- and X-ray
		images},
	\newblock \url{https://thirona.eu/cad4covid/},
	\newblock Accessed: 2020-05-14.
	
	\bibitem{StoCastiNews}
	{Business Insider - OSF HealthCare Partners with StoCastic to Use New AI System
		to Move Coronavirus Triage Out of the Hospital},
	\newblock
	\url{https://markets.businessinsider.com/news/stocks/osf-healthcare-partners-with-stocastic-to-use-new-ai-system-to-move-coronavirus-triage-out-of-the-hospital-1029036929},
	\newblock Accessed: 2020-05-14.
	
\end{thebibliography}


\begin{figure}
  \includegraphics[width=40mm]{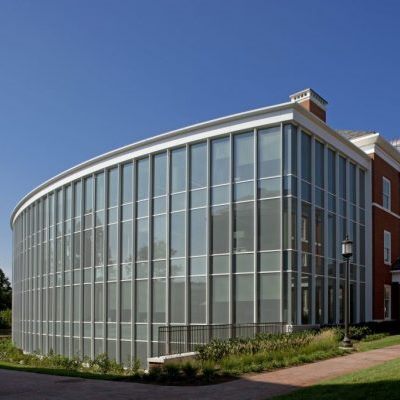}
  \caption*{The mission of The Malone Center for Engineering in Healthcare at Johns Hopkins University is to catalyze and accelerate the development of research-based innovations that advance the effectiveness and efficiency of health care.}
\end{figure}

\end{document}